\documentclass[10pt,journal,comsoc]{IEEEtran}
 
\usepackage[T1]{fontenc}
\usepackage[utf8]{inputenc}
\usepackage{cite}
\usepackage{amsmath,amssymb}
\usepackage{graphicx}
\usepackage{booktabs}
\usepackage{tabularx}
\usepackage{multirow}
\usepackage{array}
\usepackage[hyphens]{url}
\usepackage[breaklinks=true]{hyperref}

\usepackage{listings}
\usepackage{xcolor}
\usepackage{microtype}
\usepackage{tikz}
\usetikzlibrary{arrows.meta,positioning,calc}
\usepackage{balance}
\definecolor{preprintyellow}{HTML}{FFF200}

\usepackage{fancyhdr}
\fancypagestyle{preprintstyle}{%
  \fancyhf{}%
  \fancyhead[C]{\colorbox{preprintyellow}{\textcolor{black}{\footnotesize\sffamily This is the author's pre-print version. Paper under submission to an IEEE venue.}}}%
}

\begin{document}
\title{A Lightweight, Open-Source Network Digital Twin Blueprint for IoT-Edge Environments}
 
\author{Josevany~do~Amaral,~Rute~C.~Sofia%
\thanks{Josevany do Amaral is with fortiss GmbH, Munich, Germany, and ISTEC, Lisbon (e-mail: josevany.amaral@my.istec.pt).}
\thanks{Corresponding author: Rute C. Sofia (ORCID 0000-0002-7455-5872) is with fortiss GmbH, Munich, Germany (e-mail: sofia@fortiss.org).}
}
 
\pagestyle{preprintstyle}
\maketitle
\thispagestyle{preprintstyle}

\begin{abstract}
IoT-Edge-Cloud infrastructures are increasingly heterogeneous, and
experimenting across the continuum is difficult and risks disrupting
live systems. Network Digital Twins (NDT) are a pragmatic answer, yet
practitioners lack a structured account of three things: how to build
an NDT from open tools, how to model wireless impairments, and how to
understand whether the twin is faithful to the physical network, or not. This tutorial introduces NDT through the IRTF reference model and walks through a lightweight open-source blueprint built on Containerlab, Open vSwitch, \textit{Open Network Operating System (ONOS)}, and Prometheus/Grafana. It then explains how to define wireless impairment profiles (static,
model-based, trace-driven, AI-generated, and live), and what each
approach can and cannot reproduce. Finally, it presents a workflow to validate the twin against the physical network. A proof-of-concept
Edge case study illustrates the procedure, including the pitfalls encountered and the lessons learned. Directions for future research are outlined. 
\end{abstract}
 
\begin{IEEEkeywords}
Network Digital Twin, IoT-Edge-Cloud Continuum, SDN, tutorial.
\end{IEEEkeywords}

\section{Introduction}
\label{sec:intro}

The IoT-Edge-Cloud continuum spans resource-constrained sensor nodes,
wireless access networks, Edge compute clusters, and cloud backends,
creating multi-tier environments where network behavior is
heterogeneous and non-stationary. Validating configurations,
stress-testing failure recovery, and verifying \textit{Quality of
Service (QoS)} policies in such environments, without disrupting
production traffic, is a fundamental operational challenge.

\textit{Network Digital Twins (NDTs)} address this challenge by
providing a virtual replica of the physical network, one that mirrors
behavioral characteristics closely enough to serve as a safe sandbox
for experimentation. The \textit{Internet Research Task Force (IRTF)}
Network Management Research Group defines an NDT as \textit{"an
advanced platform for network emulation, serving as a tool for
scenario planning, impact analysis, and change
management"}~\cite{zhou2026irtf}. 

What distinguishes an NDT from a classical simulator or from a
standalone emulator is the mapping between the virtual and the
physical network: the twin is calibrated from measurements of the
physical network and compared with it. In this paper, the mapping is
established using measurement-calibrated profiles. Closing the loop
automatically is discussed in future work and is not evaluated.

Building such a twin raises three practical questions. First, how can
an NDT be built from open, low-cost tools that integrate
container-native topology emulation, programmable
\textit{Software Defined Network (SDN)} control, and a live
observability pipeline? Second, how can wireless impairments be
modeled in it and what can a given modeling approach reproduce? Third, how can one show that the twin is faithful to the physical network? 

This tutorial answers these three questions with a hands-on approach, built on freely available tooling, that a practitioner can replicate on modest hardware.

The tutorial is aimed at network engineers and researchers who
operate or study Edge and \textit{Internet of Things (IoT)} networks and are familiar with Linux networking and traffic control~\cite{jurgelionis2011},
containers, and basic SDN concepts~\cite{kreutz2015}. Readers new to the NDT concept can start from~\cite{almasan2022,zhou2026irtf}. After reading this tutorial, the reader should be able to (i) deploy a container-based NDT of a small wireless Edge network, (ii) define and calibrate network-layer impairment profiles from measurements, and (iii) design a comparison that supports, or refutes, a fidelity claim about the twin. The approach is illustrated on a far-Edge testbed at the fortiss research institute's IIoT
Lab\footnote{\url{https://www.fortiss.org/en/research/fortiss-labs/detail/iiot-lab}}, which hosts an IoT-Edge-Cloud testbed.

The tutorial has a deliberately limited scope. It is not a literature
survey, and the background material is limited to what the reader needs to follow the three parts. The tutorial makes the following contributions (Fig. \ref{fig:roadmap}):
\begin{enumerate}
  \item \textbf{Building the NDT:} open-source step-by-step blueprint mapped onto the IRTF reference model, together with the reusable tooling (topology manifests, SDN control logic, impairment scripts and observability stack) needed to reproduce it (Section~\ref{sec:blueprint}).
 \item \textbf{Modeling wireless impairments:} an overview of the
        ways to define impairment profiles (static, model-based,
        trace-driven, \textit{Artifical Intelligene (AI)}-generated, and live), a procedure for
        deriving static profiles from testbed measurements, and
        explicit statements of what behavioral and protocol-level
        emulation can and cannot reproduce
        (Section~\ref{sec:modeling}).
  \item \textbf{Validating fidelity:} a workflow for comparing the
        twin with the physical network, based on paired measurements,
        pre-declared exclusion rules, separate calibration and
        validation data, and distributional comparison
        (Section~\ref{sec:validation}).
  \item \textbf{Worked example and lessons learned:} a proof-of-concept
        case study, covering impairment profiles, SDN traffic engineering, QoS-based traffic prioritization, and twin vs. physical comparison (Sections~\ref{sec:case}-\ref{sec:lessons}).
\end{enumerate}

\begin{figure}[htp!]
\centering
\resizebox{\columnwidth}{!}{%
\begin{tikzpicture}[
  node distance=0.35cm and 0.5cm,
  q/.style={draw, rounded corners=2pt, align=center, text width=2.7cm,
            inner sep=3pt, font=\footnotesize, fill=gray!8},
  p/.style={draw, rounded corners=2pt, align=center, text width=2.4cm,
            inner sep=3pt, font=\footnotesize, fill=white},
  arr/.style={-{Stealth[length=5pt]}, thick}
]
\node[q] (q1) {\textbf{Q1}\\How to build an NDT?};
\node[q, below=of q1] (q2) {\textbf{Q2}\\How to model wireless impairments?};
\node[q, below=of q2] (q3) {\textbf{Q3}\\How to show the twin is faithful?};
\node[p, right=of q1] (p1) {\textbf{Blueprint}\\Sec.~III};
\node[p, right=of q2] (p2) {\textbf{Modeling}\\Sec.~IV};
\node[p, right=of q3] (p3) {\textbf{Validation}\\Sec.~V};
\foreach \i in {1,2,3} \draw[arr] (q\i) -- (p\i);
\node[p, right=0.7cm of p2, fill=gray!8] (cs) {\textbf{Case study and validation}\\Wi-Fi Edge cluster\\Sec.~VI--VIII};
\foreach \i in {1,2,3} \draw[arr] (p\i.east) -- (cs.west);
\end{tikzpicture}}
\caption{Organization of the tutorial: from practical questions to case study validation.}
\label{fig:roadmap}
\end{figure}

The remainder of the paper is organized as follows.
Section~\ref{sec:related} provides background information and positions the tutorial against related work.
Section~\ref{sec:blueprint} presents the blueprint,
Section~\ref{sec:modeling} explains the impairment modeling and
calibration, and Section~\ref{sec:validation} presents the
fidelity-validation workflow. Section~\ref{sec:case} describes the
case study, its testbed, datasets, and experiments, and
Section~\ref{sec:results} reports the results.
Section~\ref{sec:lessons} analyzes findings, limitations, and
pitfalls, and Section~\ref{sec:conclusion} concludes and outlines
future research.

\section{Related Work and Background}
\label{sec:related}

Digital twinning originated in industrial
manufacturing~\cite{grieves2017} and has since been adapted to
networking. Early emulation platforms such as GNS3 and CORE enabled
topology virtualization, but lacked integration with real-time
telemetry and SDN control planes. Mininet later established a
standard for SDN experimentation~\cite{lantz2010}, although its
kernel-space architecture limits container diversity and scale.
Containerlab~\cite{containerlab2021} addresses this with a
container-native, declarative topology engine: nodes are standard
Docker containers, each carrying its own full networking stack and
monitoring agents, described in a single \textit{Yet Another Markup Language (YAML)} manifest; deployment
and tear-down take seconds, and, unlike Mininet, no kernel patches
are needed. 

Containerlab is used as one of the key technologies for NDT. A limitation discussed in the paper is that, in Containerlab/OVS
environments, standard \textit{Maximum Transmit Unit (MTU)}-sized transfers can exhaust the \textit{Token Bucket Filter (TBF)} before packets leave the container, producing artificially low
throughput unrelated to the configured cap.

NDT validation work to date has focused mainly on datacenter
scenarios or analytical delay models~\cite{Pandey2023}; frameworks
targeting IIoT or wireless Edge are less common and typically stop
short of SDN control-plane testing or long-term distribution
matching. 
Mininet-WiFi~\cite{fontes2015} extends Mininet with virtual wireless stations and access points, and discrete-event simulators such as ns-3~\cite{riley2010} model wireless links at the PHY/MAC level. Neither provides the container-native, SDN-controlled twin with profile calibration and fidelity validation presented here.
Wireless impairment injection via \texttt{tc netem} is an established
practice~\cite{roughan2005}, but its interaction with Linux TBF shapers in containerized environments, including the segment-starvation artifact has not been previously
documented.

The IoT-Edge-Cloud continuum integrates heterogeneous cyber-physical
systems spanning far and near-Edge devices through to the Cloud,
crossing multiple administrative domains and mixing Wi-Fi, Ethernet,
and cellular technologies; a policy change at the Edge can cascade
into cloud-bound telemetry streams. Kubernetes-derived orchestration, such as K3s~\cite{k3s2023}, is increasingly deployed on the Edge, but its \textit{Container Network Interface (CNI)} overlay adds real
integration complexity where it meets SDN:
ONOS-installed flow rules can conflict with CNI-managed
\texttt{iptables} rules, and monitoring targets must be manually configured. This NDT-Kubernetes (telco-cloud) boundary is why this work
uses manual provisioning rather than layering the NDT on K3s
directly. Frameworks such as the \textit{Cognitive Decentralized Edge Cloud Orchestration (CODECO)}~\cite{sofia2026} are starting to
bridge communication and computation concerns holistically, pointing
to where this tooling gap will need to close.

The IRTF NDT reference architecture~\cite{zhou2026irtf} defines five
functional elements, \emph{data}, \emph{models}, \emph{mapping},
\emph{interfaces}, \emph{logic}, and five construction challenges:
large-scale data management, interoperability between heterogeneous
devices, data-modeling complexity, real-time synchronization, and
security. At IIoT-Edge scale, the first three are the most pressing,
and are what our blueprint is built to
address. That blueprint relies on two other building blocks
detailed there: ONOS~\cite{onos2014}, an SDN controller implementing
carrier-grade OpenFlow-based intent-driven forwarding, and the
Prometheus~\cite{prometheus2016} and Grafana stack, which provides the
data-repository and visualization functions needed for analyzing twin fidelity to the physical network.

Three open problems follow from this and shape the rest of the
article: faithful reproduction of heavy-tailed wireless RTT
distributions, which parametric \texttt{netem} models cannot capture;
dynamic recalibration as physical channel conditions change; and
federated twinning across multiple administrative
domains~\cite{zhou2026irtf}. 
\section{Building the NDT}
\label{sec:blueprint}

This section presents a generic blueprint for a container-based NDT that can be used in different scenarios, including Edge constrained scenarios. It builds the twin in three steps, considering the functional planes provided by the IRTF NDT model, using only open-source components. 

\subsection{Mapping to the IRTF Reference Model}
\begin{figure}[htp!]
  \centering
  \includegraphics[width=\columnwidth]{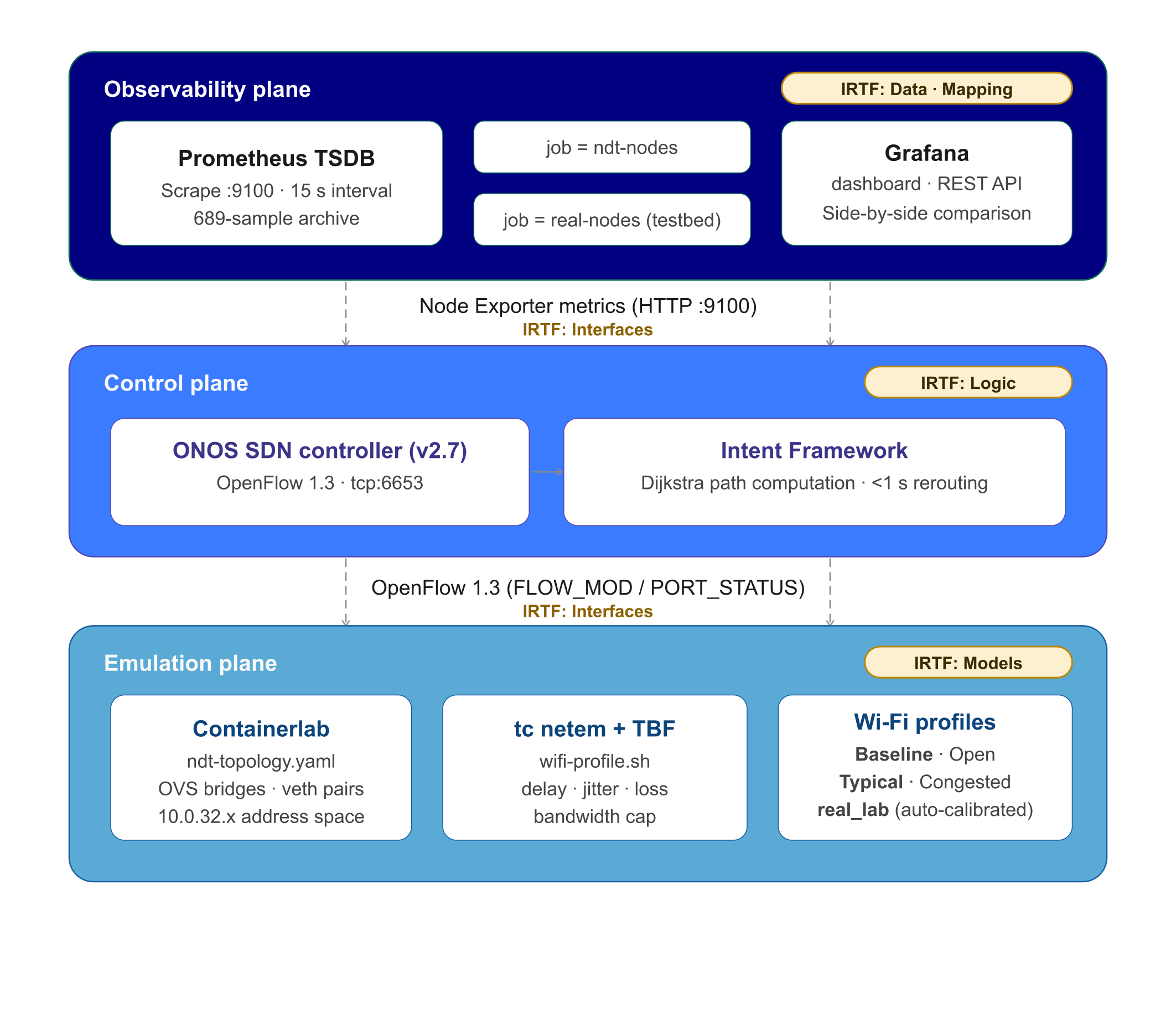}
  \caption{NDT three-plane architecture and component interactions.}
  \label{fig:arch}
\end{figure}

The blueprint NDT is structured around three functional planes
(Fig.~\ref{fig:arch}): the \textit{emulation plane} instantiates the
topology and injects channel impairments; the \textit{control plane}
provides SDN management via ONOS; and the \textit{observability
plane} collects, stores, and visualizes telemetry from both the NDT
and the physical network. Table~\ref{tab:irtfmap} shows how the five
IRTF elements are realized by the components used in the following
steps.

\begin{table}[htp!]
\centering
\caption{Realization of the IRTF NDT elements in the blueprint.}
\label{tab:irtfmap}
\small
\begin{tabular}{p{1.5cm} p{6.2cm}}
\toprule
\textbf{Element} & \textbf{Realization} \\
\midrule
Data & Prometheus time series with telemetry from the twin and the physical network \\
Models & Containerlab topology manifest and impairment profiles (Section~\ref{sec:modeling}) \\
Mapping & Shared observability plane; profiles derived from physical measurements \\
Interfaces & OpenFlow~1.3 (ONOS to OVS), HTTP scraping of Node Exporter, impairment script \\
Logic & ONOS Intent Framework (path computation and rule installation) \\
\bottomrule
\end{tabular}
\end{table}

\subsection{Step 1: Emulation Plane}

The topology is declared in a Containerlab YAML manifest, mirroring
the topology of the physical network to be twinned. Open~vSwitch
bridges replace standard Linux bridges at each wireless node,
connected to ONOS via OpenFlow~1.3 for programmatic flow control.
Impairment is applied at the \textit{Access Point (AP)}-facing interfaces via \texttt{tc
netem} (delay, jitter, loss) plus a token-bucket shaper (TBF) for the
rate cap, driven by a reusable Python script that takes an impairment
profile as input. How profiles are defined, calibrated, and updated
is the subject of Section~\ref{sec:modeling}.

Table~\ref{tab:buildsteps} provides an example of the commands that can be used to instantiate the twin.

\begin{table}[htp!]
\centering
\caption{Example commands to instantiate the twin.}
\label{tab:buildsteps}
\small
\begin{tabular}{p{0.6cm} p{2.6cm} p{3.9cm}}
\toprule
\textbf{Step} & \textbf{Action} & \textbf{Command (abridged)} \\
\midrule
1 & Create OVS bridge and attach to ONOS & \texttt{ovs-vsctl add-br ap}; \texttt{ovs-vsctl set-controller ap tcp:127.0.0.1:6653} \\
2 & Deploy topology & \texttt{containerlab deploy -t ndt.clab.yml} \\
3 & Cap the rate (root qdisc) & \texttt{tc qdisc add dev <if> root handle 1: tbf rate <R> burst <b> latency 400ms} \\
4 & Add delay, jitter, loss (child qdisc) & \texttt{tc qdisc add dev <if> parent 1:1 handle 10: netem delay <d> <j> loss <p>} \\
5 & Start telemetry & \texttt{node\_exporter} in each container; Prometheus scrape job per domain \\
\bottomrule
\end{tabular}
\end{table}

\subsection{Step 2: Control Plane}
An SDN control plane such as ONOS connects to each bridge (e.g., \textit{Open vSwitch, OVS}~\cite{pfaff2015}). During topology discovery, it installs default forwarding rules; its Intent Framework then translates path intents into OpenFlow rules via Dijkstra's shortest-path computation, pushed as flow-mod updates. When a link-state change is reported, ONOS recomputes paths and issues updated rules. A degradation that does not change the link state, such as added delay, does not trigger this process, so the twin needs an explicit detection step. Table~\ref{tab:reroute} suggests some ways to handle different degradation triggers. The reaction time should be reported in two parts: the \emph{detection time}, from the start of the degradation to the decision to reroute, and the \emph{reconfiguration time}, from that decision to the installation of the new flow rules and the traffic using the new path.

\begin{table}[htp!]
\centering
\caption{Ways to trigger rerouting after a delay degradation.}
\label{tab:reroute}
\small
\begin{tabular}{p{2.0cm} p{2.9cm} p{2.4cm}}
\toprule
\textbf{Mechanism} & \textbf{Trigger} & \textbf{Detection time} \\
\midrule
Link-state event & Port or link goes down; not raised by added delay & Immediate, but only for failures \\
Active probing & Periodic latency probe exceeds a threshold, then the intent is updated through the ONOS REST API & Probe interval \\
Telemetry alert & Prometheus alert on \textit{Round-Trip Time (RTT)} or queue metrics calls the ONOS REST API & Scrape interval plus alert delay \\
Scripted update & The experiment script updates the intent when it injects the delay & Negligible; no detection is exercised \\
\bottomrule
\end{tabular}
\end{table}

\subsection{Step 3: Observability Plane}
The observability plane has two data paths (Table~\ref{tab:obs}).
\emph{Passive telemetry} comes from Prometheus Node Exporter agents
that run in every emulated container via Containerlab \texttt{exec}
directives, exposing per-interface byte/packet counters, CPU, memory,
and load. Prometheus scrapes two jobs at a fixed interval (e.g., every
15 seconds): the emulated NDT containers and the nodes of the physical
network (access point and wireless clients). \emph{Active measurements}
can be produced by tools such as \texttt{ping} and \texttt{iperf3} runs, whose output a collection script parses and stores with timestamps; these provide the RTT, jitter, loss, and throughput values used for the fidelity
comparison.
The NDT and physical network feed the same Prometheus instance and Grafana
dashboards, and use the same metric names with a label identifying the
specific domain (e.g., twin or physical), to allow for a visual comparison of behavior side by side. For paired measurements, the clocks of both domains must be synchronized (e.g., with the \textit{Network Time Protocol (NTP)}) so that runs can be
aligned in time.

\begin{table}[htp!]
\centering
\caption{Data paths of the observability plane.}
\label{tab:obs}
\small
\begin{tabular}{p{1.7cm} p{2.5cm} p{3.1cm}}
\toprule
\textbf{Path} & \textbf{Source} & \textbf{Use} \\
\midrule
Passive telemetry & Node Exporter, scraped by Prometheus & Interface counters, CPU, memory, load; live dashboards \\
Active measurements & \texttt{ping}, \texttt{iperf3}, parsed by a script & RTT, jitter, loss, throughput; fidelity comparison \\
\bottomrule
\end{tabular}
\end{table}

\section{Modeling and Calibrating Wireless Impairments}
\label{sec:modeling}
\subsection{What the Model Must Reproduce}
Wi-Fi links exhibit delay and jitter, bursty and interference-correlated loss, rate adaptation, airtime sharing, and occasional multi-second latency spikes. The model must decide which of these to reproduce and at which layer. Two levels are available in a container-based twin (Table~\ref{tab:levels}).
\emph{Behavioral emulation} imposes the end-to-end effects
of the wireless link on ordinary Ethernet traffic. \emph{Protocol level
emulation} runs the real IEEE~802.11 stack over a simulated
radio medium. The choice depends on what the twinned application
observes: workloads that only see delay, loss, and throughput can be served via behavioral emulation; studies of association, rate adaptation, or frame formats need protocol emulation support as well.

\begin{table}[htp!]
\centering
\caption{Two levels of Wi-Fi emulation in a container-based twin.}
\label{tab:levels}
\small
\begin{tabular}{p{1.5cm} p{2.9cm} p{2.9cm}}
\toprule
 & \textbf{Behavioral} & \textbf{Protocol} \\
\midrule
Mechanism & \texttt{tc netem} and TBF on \texttt{veth} & \texttt{mac80211\_hwsim} with \texttt{wmediumd} \\
Frames & Ethernet (802.3) & Real 802.11 (radiotap) \\
Reproduces & Delay, jitter, loss, bursts, reordering, rate cap & Association, WPA2/3, MCS rate selection, per-station SNR, hidden node, power save, WMM \\
Does not reproduce & MAC contention, rate adaptation, hidden node, power save, RF interference & Multipath, OFDM/OFDMA PHY, external interference, driver quirks \\
Setup cost & Low; no kernel module & Higher; host kernel module and radio setup \\
\bottomrule
\end{tabular}
\end{table}

\subsection{Behavioral Level Emulation}
Injection of wireless impairments through \texttt{tc netem} is an established practice~\cite{roughan2005}. The blueprint applies delay, jitter, and loss with \texttt{netem} plus a token-bucket shaper for the rate cap. Table~\ref{tab:netem} maps each wireless characteristic to the
mechanism that reproduces it. Bursty loss uses the Gilbert-Elliott
model~\cite{gilbert1960,elliott1963} available as \texttt{netem loss gemodel}; frame corruption and
reordering are also available.

\begin{table}[htp!]
\centering
\caption{Wireless characteristics and their behavioral level mechanism.}
\label{tab:netem}
\small
\begin{tabular}{p{2.6cm} p{4.6cm}}
\toprule
\textbf{Characteristic} & \textbf{Mechanism} \\
\midrule
Latency, jitter & \texttt{netem delay <d> <j>} \\
Random loss & \texttt{netem loss <p>} \\
Burst loss & \texttt{netem loss gemodel} (two-state Markov) \\
Corruption, reordering & \texttt{netem corrupt}, \texttt{netem reorder} \\
Bandwidth cap & \texttt{tbf rate <R> burst <b>} \\
\bottomrule
\end{tabular}
\end{table}

The behavioral emulation therefore reproduces link-level effects as seen by the network layer. However, it does not model channel contention, retransmissions, rate adaptation, mobility, or hidden-terminal effects, which require PHY/MAC-aware emulation or simulation. 

\subsection{Protocol Level Emulation}
When the application depends on the behavior of IEEE 802.11 itself, the Linux \texttt{mac80211\_hwsim}~\cite{hwsim} driver creates virtual radios that use the real \texttt{mac80211} stack and \texttt{wmediumd}~\cite{wmediumd} simulates the
shared medium. A static matrix of SNR values per pair determines
reception and, through it, the rate that is selected; an access point
(\texttt{hostapd}) and stations (\texttt{wpa\_supplicant}) run in the
containers. Distance and obstruction are expressed by lowering the
SNR between nodes, and asymmetric values reproduce the hidden-node
problem. The packet captures show genuine 802.11 frames.

Protocol level emulation still has limits. Because \texttt{wmediumd} uses a static SNR matrix, it does not model multipath propagation, time-varying interference, or the OFDM/OFDMA physical layer, and the supported 802.11 amendments depend on the driver and \texttt{wmediumd} version.

\subsection{Defining Impairment Profiles}
\label{subsec:profiles}
\begin{figure}[htp!]
\centering
\resizebox{\columnwidth}{!}{%
\begin{tikzpicture}[
 node distance=0.4cm,
 src/.style={draw, rounded corners=2pt, align=center, text width=2.3cm,
        inner sep=2pt, font=\scriptsize, fill=gray!8},
  box/.style={draw, rounded corners=2pt, align=center, text width=2.4cm,
              inner sep=3pt, font=\footnotesize, fill=white},
  arr/.style={-{Stealth[length=5pt]}, thick}
]
\node[src] (s1) {Static};
\node[src, below=0.15cm of s1] (s2) {Model-based};
\node[src, below=0.15cm of s2] (s3) {Trace-driven};
\node[src, below=0.15cm of s3] (s4) {AI-generated};
\node[src, below=0.15cm of s4] (s5) {Live (closed loop)};
\node[box, right=1.0cm of s3] (prof) {\textbf{Profile}\\delay, jitter,\\loss, rate cap};
\node[box, right=0.8cm of prof] (imp) {\textbf{Impairment}\\\texttt{netem} + TBF\\on AP-facing\\interface};
\foreach \i in {1,...,5} \draw[arr] (s\i.east) -- (prof.west);
\draw[arr] (prof) -- node[above, font=\scriptsize]{script} (imp);
\end{tikzpicture}}
\caption{Five ways to define a wireless impairment profile.}
\label{fig:profiles}
\end{figure}

A \emph{profile} (Fig.~\ref{fig:profiles}) is a named set of behavioral emulation parameters (delay, jitter, loss, rate cap) that the impairment script applies to an interface. Profiles can be defined in multiple ways as proposed in Table~\ref{tab:profdef}.

\begin{table}[htp!]
\centering
\caption{Ways to define an impairment profile.}
\label{tab:profdef}
\small
\begin{tabular}{p{1.4cm} p{2.9cm} p{2.9cm}}
\toprule
\textbf{Definition} & \textbf{What it captures} & \textbf{Main limitation} \\
\midrule
Static & One operating point (e.g., median values) & No variation over time \\
Model-based & Distribution, correlation, or bursts, from a few parameters & Model may not match the channel \\
Trace-driven & Measured time series replayed on the twin & Valid only for the recorded conditions \\
AI-generated & Variation learned from data, or conditions not measured & Depends on training data; needs validation \\
Live (closed loop) & Current state of the physical network & Needs a running physical network\\
\bottomrule
\end{tabular}
\end{table}

\paragraph{Static profiles}
A static profile fixes one value per parameter. This paper uses four,
Baseline, Open, Typical and Congested, derived from testbed
measurements and covering increasing congestion. Each profile is derived in three steps:
\begin{enumerate}
 \item \textbf{Measure} the specific metric, e.g., RTT, jitter, and loss with \texttt{ping}, and throughput under the condition to be
       represented.
 \item \textbf{Summarize} each metric over equal-length windows, using
        the median (and the interquartile range for jitter).
 \item \textbf{Map} the summaries to parameters: delay is half the
        median RTT in each direction, jitter is the interquartile
        range, loss is the observed rate, and the rate cap is the
        median throughput.
\end{enumerate}
Static profiles are simple and repeatable, but reproduce only
one operating point. Conditions not measured in the lab, such as a
link at the edge of coverage, can be added as \emph{nominal} profiles,
labeled as such because their values are typical figures and not
measurements of the twinned network.

\paragraph{Model-based profiles} A model generates the impairment from a few parameters fitted to an initial dataset, so variation can be reproduced without replaying a full measurement trace. Two time scales matter.
\emph{Per-packet variation} uses \texttt{netem} options:
\begin{itemize}
  \item \textbf{Delay distribution:} normal, Pareto, or a table
        measured from data, in place of uniform jitter, with a
        correlation parameter so that consecutive packets have
        similar delays.
  \item \textbf{Correlated random loss.}
  \item \textbf{Gilbert-Elliott loss} (\texttt{loss gemodel}): the
        channel moves between a good and a bad state, and the loss is higher in the bad state. The transitions happen per packet,
        so this produces bursts of a few lost packets as caused by
        interference.
\end{itemize}

\paragraph{Trace-driven profiles} A trace-driven profile replays a measured time series of delay, loss, and throughput. The script reads one sample per interval and updates the \texttt{netem} and shaper parameters accordingly. This keeps the real temporal pattern of one recorded period. Its limits are that it is valid only for the recorded conditions and for the traffic that produced them, so it does not generalize to other loads.

\paragraph{Live profiles} A live profile is generated from current
measurements of the physical network instead of recorded or fixed
values. It is designed so that, as the channel degrades, the next
profile injects more delay, jitter, and loss, and less bandwidth, so
that the twin follows the physical network.
 
\paragraph{AI-generated profiles} A learned model can also produce profiles. AI profiles can be derived via trace generation or parameter prediction. In trace generation, a model trained on measured traces (e.g., a generative time-series model) produces new synthetic delay, loss, and throughput sequences with similar statistics, which are then replayed as a trace-driven profile. This extends a short recording to longer or more varied runs. In parameter prediction a model maps a description of the conditions (e.g., number of clients, signal strength, offered load) to profile parameters so that conditions not measured in the lab can be interpolated.

AI-generated profiles need to be treated like any other profile: the
parameters are bounded to valid ranges, the model version and random
seed are recorded for repeatability, and the result is compared with
the physical network using the validation workflow of
Section~\ref{sec:validation}.
\section{Validating Fidelity}
\label{sec:validation}
 
A twin is only useful if its fidelity to the physical network is
demonstrated. Fidelity is understood here as the degree to which the twin reproduces the distribution of the metrics of interest, measured under comparable conditions, within tolerances that the user defines for the intended use of the twin. The general steps are represented in Fig. \ref{fig:valworkflow}.
 
\begin{figure}[htp!]
\centering
\resizebox{\columnwidth}{!}{%
\begin{tikzpicture}[
  node distance=0.45cm,
  stepbox/.style={draw, rounded corners=2pt, align=left, text width=6.1cm,
                  inner sep=3pt, font=\footnotesize, fill=gray!8},
  num/.style={circle, draw, fill=white, inner sep=1.5pt, font=\scriptsize\bfseries},
  arr/.style={-{Stealth[length=5pt]}, thick}
]
\node[stepbox] (s1) {\textbf{Define fidelity targets}\\
  metrics (RTT, jitter, loss, throughput), statistic (median, P95), tolerance};
\node[stepbox, below=of s1] (s2) {\textbf{Design a paired experiment}\\
  same command, same time, both networks, equal windows; new data, not used to build the profile};
\node[stepbox, below=of s2] (s3) {\textbf{Declare exclusion rules in advance}\\
  failed runs defined before collection; same rule in both networks};
\node[stepbox, below=of s3] (s4) {\textbf{Collect with controlled commands}\\
  one versioned script; log command and raw output per row};
\node[stepbox, below=of s4] (s5) {\textbf{Compare distributions and report}\\
  percentile error, CDF overlay, sample counts; is each tolerance met?};
\foreach \a/\b in {s1/s2,s2/s3,s3/s4,s4/s5}
  \draw[arr] (\a.south) -- (\b.north);
\foreach \i in {1,...,5}
  \node[num, anchor=east] at ([xshift=-2pt]s\i.west) {\i};
\draw[arr, dashed] (s5.east) -- ++(0.5,0) |- node[pos=0.25, right, font=\scriptsize, align=left]
  {not met:\\revise\\profile} (s1.east);
\end{tikzpicture}}
\caption{Validation workflow: five steps from fidelity targets to a
tolerance-based decision.}
\label{fig:valworkflow}
\end{figure}

The validation workflow steps are as follows:
\begin{enumerate}
  \item \textbf{Define fidelity targets:} choose the metrics that matter
        for the intended use (e.g., RTT, jitter, loss, throughput) and
        a tolerance for each, expressed as an acceptable error in a
        chosen statistic (e.g., median, P95).
  \item \textbf{Design a paired experiment:} run the same test, with the
        same command, at the same time on the physical network and on
        the twin, over equal observation windows.
  \item \textbf{Declare exclusion rules in advance:} define which runs
        count as failed (e.g., server unreachable, timeout), before any
        data are collected, and apply the rule identically to both
        environments.
  \item \textbf{Collect with controlled commands:} execute measurements
        from a single versioned script and log the exact command and
        raw output with each row.
  \item \textbf{Compare distributions and report:} compare per-metric
        distributions, report sample counts
        and results, and state whether each
        tolerance is met.
\end{enumerate}
 
\subsection{Principles}
Five principles underlie the workflow:
\begin{itemize}
  \item \textbf{Paired, simultaneous measurement:} both datasets cover
        the same period, so changes in the physical network over
        time (time of day, competing traffic) do not bias the
        comparison. With static, bimodal, or trace-driven profiles
        the twin does not see the same channel realization, so
        runs are compared as distributions only.
  \item \textbf{Pre-declared exclusion rules:} excluding runs after the
        fact, or in only one environment, can bias the comparison;
        counts of excluded runs must be reported.
  \item \textbf{Equal observation windows} and reported sample counts.
  \item \textbf{Distributional comparison}, not only means and medians,
        since the physical channel is heavy-tailed and means are
        sensitive to a few outliers.
  \item \textbf{Separate calibration and validation data:} the
        measurements used to derive or fit a profile must not be
        used to validate it. Collect a new and independent dataset for comparison.
\end{itemize}
 
\subsection{Metrics and Comparison Methods}
\label{subsec:metrics}
Table~\ref{tab:valmetrics} summarizes what to report per metric. The
main quantity is the \emph{percentile error}, the relative difference
between the twin and the physical network at a given percentile $p$
(e.g., the median or the 95th percentile):
\begin{equation}
  E_p = \frac{Q_p^{\text{NDT}} - Q_p^{\text{NET}}}{Q_p^{\text{NET}}},
\end{equation}
where $Q_p$ is the value of the metric at percentile $p$. A small
$|E_p|$ across several percentiles indicates that the twin reproduces
both typical and tail behavior. To compare whole distributions, overlay
the empirical CDFs of the two environments and report the largest
vertical gap between them.

\begin{table}[htp!]
\centering
\caption{What to report per metric when validating a twin.}
\label{tab:valmetrics}
\small
\begin{tabular}{p{1.5cm} p{2.9cm} p{3.1cm}}
\toprule
\textbf{Metric} & \textbf{Statistics} & \textbf{Comparison} \\
\midrule
RTT & median, P95, P99 & CDF overlay, percentile error \\
Jitter & median, P95 & CDF overlay, percentile error \\
Loss & mean, share of zero-loss runs, mean burst length & distribution of per-run loss \\
UDP throughput & median, P5, P95 & CDF overlay, percentile error \\
TCP goodput & median, P5, P95 & CDF overlay, percentile error, effect of any workaround \\
\bottomrule
\end{tabular}
\end{table}

\section{Case Study: an NDT for Wi-Fi Edge Cluster Experimentation}
\label{sec:case}
 
This section applies the blueprint (Section~\ref{sec:blueprint}), the
profiles (Section~\ref{sec:modeling}), and the validation workflow
(Section~\ref{sec:validation}) to a specific Edge case study. The case study is a proof of concept: it shows how the blueprint,
profiles, and workflow are applied on one testbed.
It follows an illustrative manufacturing shop-floor scenario. On a
production line, machine controllers, sensors, and a quality-inspection
application run on Edge nodes that communicate over a shared Wi-Fi
network. The plant network operator must add a new data-intensive
application, for example a camera stream for inspection, and needs to
know beforehand whether the line's time-critical control traffic will
still be delivered when the channel becomes congested. Testing on the
live line would risk stopping production, so the operator wants a twin
that behaves like the real network and on which congestion, rerouting,
and prioritization can be tested safely. The Raspberry~Pi cluster of
this section would represent a line of Edge nodes. The scenario leads
to three questions, which five experiments shall answer: how the network
behaves under Wi-Fi impairment (E1, E2), what the SDN control plane
can do about it (E3, E5), and whether the twin can be trusted for these
decisions (E4).

\subsection{Experimental Environment}
The experimental environment considers the physical network and its twin as summarized in Table~\ref{tab:env}.

\begin{table}[htp!]
\centering
\caption{Physical network and twin.}
\label{tab:env}
\small
\begin{tabular}{p{1.4cm} p{2.9cm} p{2.9cm}}
\toprule
 & \textbf{Physical network} & \textbf{Twin} \\
\midrule
Host & Five Raspberry~Pi (ARM) & One machine, Ubuntu~22.04~LTS \\
Nodes & Kubernetes CODECO workers and a control plane node & Containers (Containerlab) \\
Network & Shared WLAN & OVS bridges with \texttt{netem} and TBF impairment \\
Control & Kubernetes/CODECO, OCM & ONOS v2.7, OpenFlow~1.3 \\
Monitoring & Node Exporter, Prometheus, Grafana & same instances, shared \\
\bottomrule
\end{tabular}
\end{table}

\begin{figure}[htp!]
  \centering
  \includegraphics[width=\columnwidth]{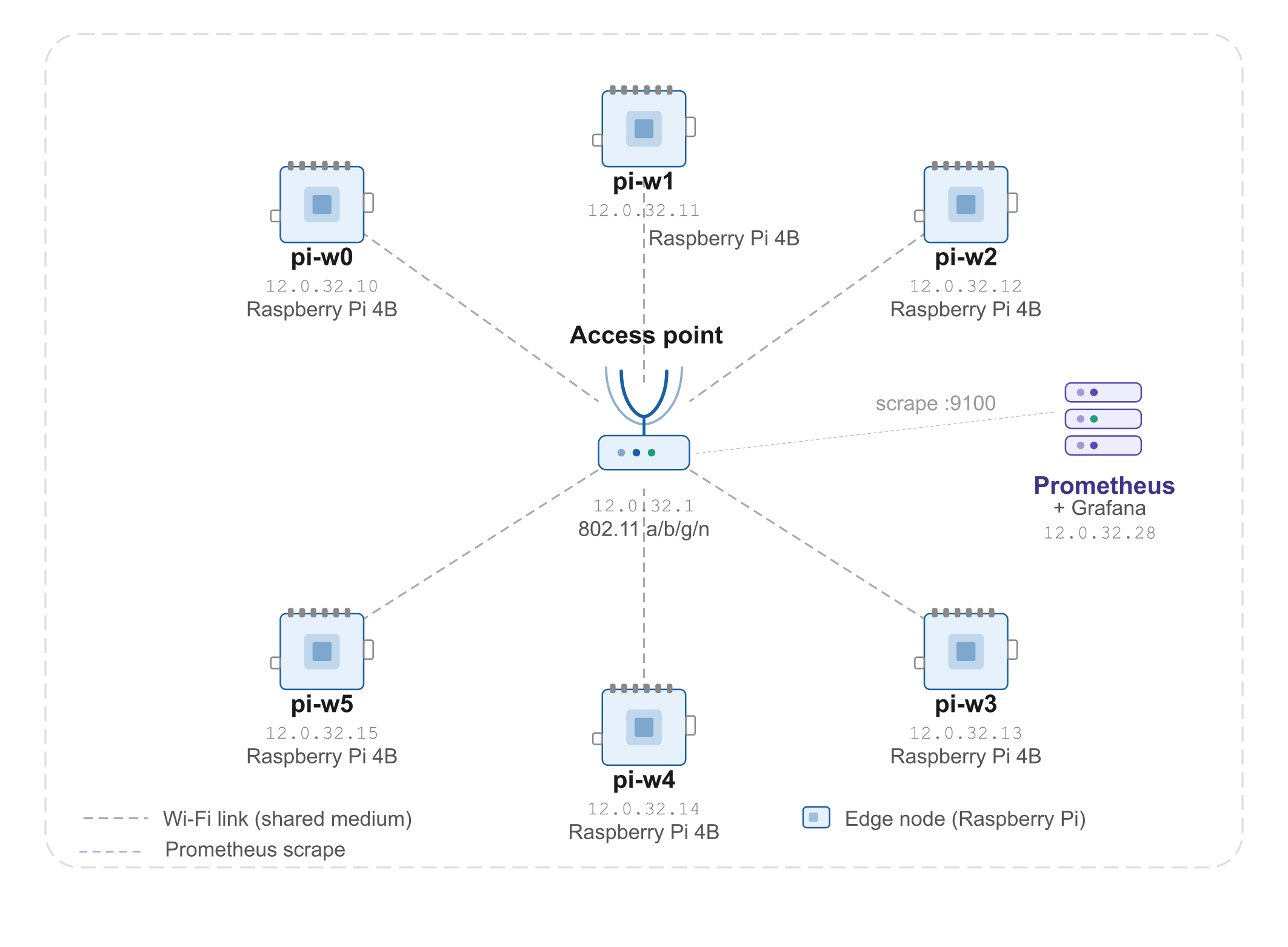}
  \caption{Reference testbed topology.}
  \label{fig:codeco}
\end{figure}

The physical testbed is part of the fortiss IIoT Lab. A multi-cluster
federated environment based on the Kubernetes CODECO extensions and
Open Cluster Management (OCM). The experiments use a cluster of
five Raspberry~Pi nodes (Fig.~\ref{fig:codeco}).

The twin is built as in Section~\ref{sec:blueprint} on a dedicated
machine with Docker, Containerlab, and ONOS/OVS. Its containers
mirror the nodes of the physical cluster, the wireless link is modeled by recurring to wireless impairment profiles, and ONOS
runs as a local process connected to the OVS bridges. Prometheus and
Grafana are shared with the physical network, so both are viewed in
the same dashboards.
 
Each node has a node exporter, with metrics scraped by a Prometheus instance. Grafana dashboards provide real-time visualization of throughput, CPU utilization, and interface statistics in both physical and emulated environments. The WLAN is not isolated, and therefore, the channel is subject to usual variability derived from network use, obstacles, channel interference. 
 
\subsection{Datasets and Test Commands}
For the case study provided in this paper, two datasets were collected, one in each environment~\footnote{Datasets and scripts are available at: \url{https://git.fortiss.org/iiot_external/experimentation-papers-replicability/ndt-paper-2026}}. The \textbf{NET} dataset was collected on the physical network (629 sessions, 2026-05-19 to 2026-05-26, after exclusions). The \textbf{NDT} dataset was collected on the twin (70 sessions, 2026-06-02 to 2026-06-03). A \emph{session} is one execution of the three commands in Table~\ref{tab:commands}; it yields one TCP, one UDP, one jitter and one loss value, and up to 20 RTT samples.

\begin{table}[htp!]
\centering
\caption{Test commands used for all measurements.}
\label{tab:commands}
\small
\begin{tabular}{p{1.6cm} p{4.6cm} p{1.2cm}}
\toprule
\textbf{Test} & \textbf{Command} & \textbf{Purpose} \\
\midrule
Latency/loss & \texttt{ping -c 20 <target>} & RTT, loss \\
TCP & \texttt{iperf3 -c <target> -t 20 -M 500} & Goodput \\
UDP & \texttt{iperf3 -c <target> -u -b 8M -t 20 -l 500} & Throughput, loss, jitter \\
\bottomrule
\end{tabular}
\end{table}

The raw data was subject to exclusions as
listed in Table~\ref{tab:exclusions}.

\begin{table}[htp!]
\centering
\caption{Excluded or missing entries by dataset and cause.}
\label{tab:exclusions}
\small
\begin{tabular}{p{5.0cm} p{2.2cm}}
\toprule
\textbf{Dataset and observation, with cause} & \textbf{Entries} \\
\midrule
\textbf{NET} Corrupt block from 14:00 on 2026-05-26: UDP up to
645\,Mbps (8\,Mbps offered), TCP up to 767\,Mbps (usual range
1--48\,Mbps), jitter up to 184\,ms (23.8\,ms maximum on other days).
Cause: the \texttt{-M 500} option was removed from the TCP command.
All rows of the block are excluded, including ping.
& 1,323 rows (60 sessions) \\
\textbf{NET} No value recorded (empty field); cause not identified.
& 53 TCP; 16 UDP, jitter, loss \\
\textbf{NDT} TCP values of at least 10\,Mbps (highest valid reading
4.92\,Mbps). Cause: column-parsing error in the collection script
after a change of the iPerf3 unit.
& 7 TCP \\
\textbf{NDT} TCP reported 0\,Mbps while UDP and ping in the same
session succeeded; cause not identified.
& 8 TCP \\
\textbf{NDT} UDP reported 100\% loss (0\,Mbps) while TCP and ping
succeeded.
& 5 sessions (UDP, jitter, loss) \\
\bottomrule
\end{tabular}
\end{table}

After these steps, NET has 576 sessions with a valid TCP value and 613
with valid UDP, jitter, and loss values. NDT has 55 and 65 sessions,
respectively.

\subsection{Validation Experiments}
To assess how far a digital twin can support the use case, five
experiments were carried out (Table~\ref{tab:scenarios}). They are grouped by the three questions of the user story. E1, E2, and E4 use the WLAN topology; E3 and E5 use the SDN topology, whose baseline throughput is higher.

\begin{table}[htp!]
\centering
\caption{Experiments, grouped by the question they answer.}
\label{tab:scenarios}
\small
\begin{tabular}{p{4.2cm} p{3.0cm}}
\toprule
\textbf{Experiment and setup} & \textbf{Metrics} \\
\midrule
\multicolumn{2}{l}{\emph{How does the network behave under Wi-Fi impairment?}} \\
\textbf{E1 Baseline:} no impairment, no \texttt{tc} (forwarding ceiling) & RTT, throughput \\
\textbf{E2 Congestion:} Congested profile & TCP/UDP throughput, loss, RTT \\
\midrule
\multicolumn{2}{l}{\emph{What can the SDN control plane do about it?}} \\
\textbf{E3 Rerouting:} 100\,ms added delay on the primary link triggers an OpenFlow intent that moves traffic to a secondary path & Latency and throughput before/after, detection and reconfiguration time \\
\textbf{E5 Prioritization:} OpenFlow priority rules; high-priority flows get dedicated resources, best-effort flows share the rest & High-priority vs.\ best-effort throughput and latency \\
\midrule
\multicolumn{2}{l}{\emph{Can the twin be trusted for these decisions?}} \\
\textbf{E4 Twin validation:} bimodal profile cycling between low and high congestion & Percentile error, CDF overlay \\
\bottomrule
\end{tabular}
\end{table}

\section{Case Study Evaluation Results}
\label{sec:results}
This section reports results for each proposed experimental scenario.
The results are organized by the three questions of the user story
(Table~\ref{tab:scenarios}).
 \begin{table*}[htp!]
  \centering
  \caption{Wi-Fi impairment profiles: injected parameters and measured NDT response.}
  \label{tab:profiles_combined}
  \renewcommand{\arraystretch}{1.15}
  \begin{tabular}{l cccc cccc}
    \toprule
    & \multicolumn{4}{c}{\textbf{Injected Impairment}} & \multicolumn{4}{c}{\textbf{Measured NDT Response}} \\
    \cmidrule(lr){2-5} \cmidrule(lr){6-9}
    \textbf{Profile} & \textbf{Delay} & \textbf{Jitter} & \textbf{Loss} & \textbf{BW Cap}
      & \textbf{RTT avg} & \textbf{TCP} & \textbf{Jitter} & \textbf{UDP Loss} \\
    \midrule
    Baseline  & 0\,ms  & 0\,ms   & 0\%    & None        & 0.079\,ms  & 2.01\,Gbps & 0.006\,ms & 0\%     \\
    Open      & 2\,ms  & 0.5\,ms & 0.05\% & 150\,Mbit/s & 4.169\,ms  & 46.8\,Mbps & 0.401\,ms & 0.049\% \\
    Typical   & 8\,ms  & 2\,ms   & 0.5\%  & 54\,Mbit/s  & 16.327\,ms & 3.98\,Mbps & 0.955\,ms & 0.5\%   \\
    Congested & 20\,ms & 6\,ms   & 3\%    & 10\,Mbit/s  & 39.920\,ms & 471\,Kbps  & 3.560\,ms & 5.9\%   \\
    \bottomrule
  \end{tabular}
\end{table*}

\subsection{How Does the Network Behave Under Wi-Fi Impairment?}
\paragraph{E1: baseline}
Table~\ref{tab:profiles_combined} shows the twin response to the
four static profiles on the WLAN topology: the left columns provide the
injected parameters while the right columns provide the measured response. Under Baseline, 2.01\,Gbps TCP throughput reflects raw OVS forwarding capacity on the single physical host of this topology, and the sub-millisecond RTT shows minimal queueing when idle. This is the upper bound for the WLAN topology; the separate SDN topology of E3 has its own higher ceiling.

\paragraph{E2: congestion} From Open to Typical to Congested, all four measured metrics degrade monotonically, as designed. The measured RTT is about twice the injected delay, since the delay is applied in each direction, and the UDP loss matches the injected loss for Open and Typical. For Congested, the measured UDP loss (5.9\%) is higher than the injected 3\%. A likely cause is queue overflow at the 10\,Mbit/s cap, where the 8\,Mbit/s UDP stream plus headers and the 6\,ms jitter leave little margin; this was not verified.
 
The TCP throughput is below the cap of each profile and the gap grows with injected loss. This is consistent with loss-limited TCP with small
segments. A standard estimate of TCP throughput under random loss,
$\mathrm{MSS}/\mathrm{RTT}\cdot 1.22/\sqrt{p}$~\cite{mathis1997},
with $\mathrm{MSS}=500$\,bytes (the \texttt{-M 500} setting) and the
measured RTT and loss, gives 52, 4.2, and 0.71\,Mbps for Open, Typical,
and Congested, compared to the measured 46.8, 3.98, and 0.471\,Mbps. The
estimate is approximate, but the RTT and loss with the small segment size are sufficient to explain the values, so no congestion-window collapse needs to be assumed. Therefore, these TCP values depend on the segment size used for the measurement (Section~\ref{sec:lessons}).

\subsection{What Can the SDN Control Plane Do About It?}
\paragraph{E3: rerouting} E3 runs on a separate SDN topology, so its throughput is compared to its own baseline of $\sim$9.14\,Gbps, not to the 2.01\,Gbps WLAN ceiling. A 100\,ms delay was injected on the primary inter-router link.
ONOS then installed updated OpenFlow rules on a secondary path in less than one second, with zero packet loss during the transition. The latency dropped from $\sim$101\,ms to $\sim$0.04\,ms, and TCP throughput
recovered from $\sim$330\,Mbps to $\sim$9.50\,Gbps. The control plane
can therefore move traffic to a secondary path, and the twin reproduces
the effect on latency and throughput. This corroborates that the ONOS Intent Framework correctly exercises Dijkstra-based path selection within the NDT's control plane.

\paragraph{E5: prioritization}
Under concurrent best-effort and high-priority flows sharing the same congested link, ONOS-installed OpenFlow priority rules successfully isolate latency-sensitive telemetry traffic from background load: high-priority flows retain sub-10\,ms RTT throughout the congestion window, while best-effort traffic absorbs the induced queueing delay.
This confirms that class-based QoS enforcement operates correctly within the NDT's control plane. However, despite consistency achieved within the single-run, short-duration scope of this initial validation, per-class throughput fairness under sustained multi-hour load remains
untested and is noted as a direction for future work
(Section~\ref{sec:conclusion}).

\subsection{Can the Twin Be Trusted for These Decisions?}
\begin{figure*}[htp!]
  \centering
  \includegraphics[width=0.75\textwidth]{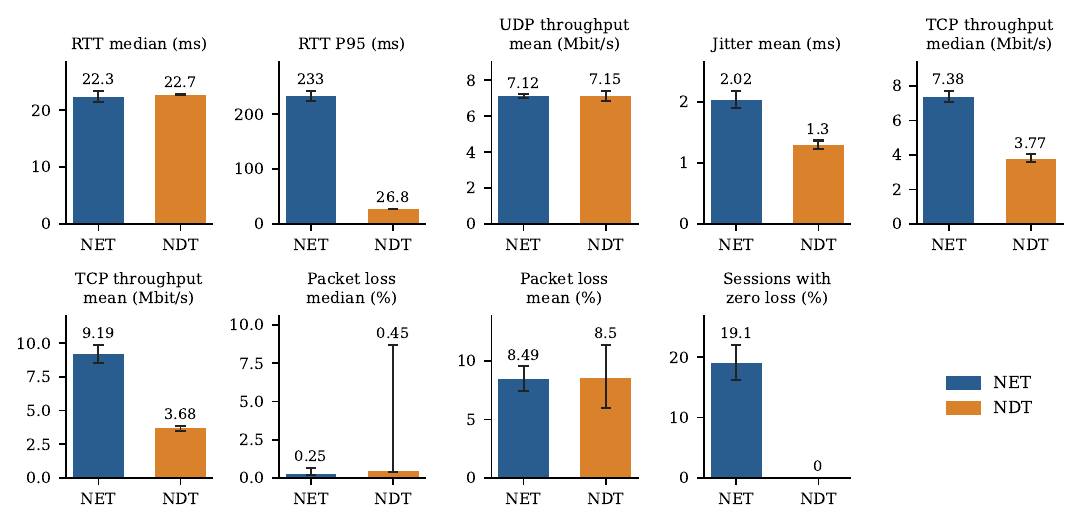}
  \caption{NET versus NDT: median, 95th percentile (P95), or mean of
  each metric, with 95\% confidence intervals.}
  \label{fig:fidelity}
\end{figure*}

\begin{figure*}[htp!]
  \centering
  \includegraphics[width=0.75\textwidth]{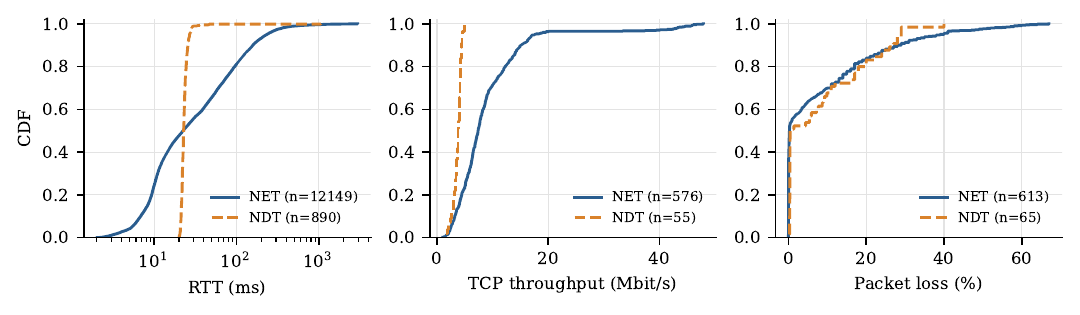}
  \caption{CDFs of RTT (logarithmic axis), TCP throughput, and packet
  loss for NET and NDT; $n$ is the number of samples (RTT) or sessions
  (TCP, loss) after exclusions.}
  \label{fig:cdf}
\end{figure*}

\paragraph{E4: NDT versus NET}
Fig.~\ref{fig:fidelity} compares the NET and NDT datasets in nine
statistics of the five metrics, and Fig.~\ref{fig:cdf} shows their
distributions. All NDT measurements use the bimodal profile
\texttt{wifi-profile-hybrid.sh}, which alternates between a calm and a
congested state.
 
The RTT median is 22.3\,ms (NET) and 22.7\,ms (NDT), the UDP
throughput mean 7.12 and 7.15\,Mbps, and the jitter mean 2.02 and
1.30\,ms. The packet loss mean is 8.49\% and 8.50\%, and the median
0.25\% and 0.45\%. The mean TCP throughput is 9.19 and 3.68\,Mbps
(a 60\% difference), and the median 7.38 and 3.77\,Mbps (49\%).
 
The NDT therefore closely reproduces the central values of RTT ($+$1.8\%) and UDP throughput ($+$0.4\%) and the mean packet loss. The jitter is 36\% lower in the NDT, and TCP throughput is 49-60\% lower. The mean
loss agreement depends on excluding the five NDT sessions with 100\%
UDP loss (Table~\ref{tab:exclusions}); with them included, the NDT
mean would be 15.0\%. The largest difference is in TCP throughput,
which is analyzed in Section~\ref{sec:lessons}.
 
\section{Findings, Limitations, and Pitfalls}
\label{sec:lessons}
\subsection{Convergence Analysis}
The NDT converges closely with the physical network in most metrics.
UDP throughput is a good example: the NET and NDT means differ by only
0.4\% (7.12 vs.\ 7.15\,Mbps). The RTT median shows a similar pattern
(22.3 vs.\ 22.7\,ms, a difference of 1.8\%), and this holds across the
full multi-day NET archive, not just under favorable short-term
conditions, although the NET and NDT datasets cover different periods.

Fig.~\ref{fig:cdf} compares the empirical distributions of RTT, TCP
throughput, and packet loss in the NET and NDT datasets. For RTT, 90\%
of the NDT values lie between 20.8 and 26.8\,ms (5th to 95th
percentile), while the NET values lie between 5.4 and 232.6\,ms, and
48\% of the NET values are below 19.9\,ms, the minimum of the NDT. For
TCP throughput, the NDT stays between 1.6 and 4.9\,Mbps, while 76\% of
the NET sessions are above 5\,Mbps. The loss distributions are close
except at zero: 19\% of the NET sessions have no loss, and the lowest
NDT loss is 0.29\%. These differences are discussed next.

\textbf{TCP Throughput Gap.} NDT TCP throughput is 60\% below the NET
mean (3.68 vs.\ 9.19\,Mbps) and 49\% below the median. The NDT stays
between 1.6 and 4.9\,Mbps regardless of the loss level, while the NET
throughput reaches 48\,Mbps, so the gap is not explained by loss alone.
The 500-byte MSS used to avoid TBF starvation limits TCP throughput,
but the Open profile reaches 46.8\,Mbps with the same setting. Sizing
the shaper burst from the configured rate (Section~\ref{sec:blueprint}),
or using HTB or HFSC, removes the need for the MSS reduction and is
worth exploring.

\textbf{Token-Bucket Starvation.} A token-bucket shaper (TBF) adds
tokens at the configured rate in steps of $1/\mathrm{HZ}$ seconds. If
the burst is smaller than the tokens added per step or smaller than a
full-size segment, packets cannot be sent, and TCP starves. Reducing the
segment size hides the problem but changes protocol behavior, so TCP
results then characterize the workaround and not the twin (Section~\ref{sec:validation}). The same \texttt{-M 500} setting is
used in both datasets (Table~\ref{tab:commands}), so the NET and NDT
TCP values are comparable to each other.

\textbf{RTT Spike Volatility.} The NET channel occasionally spikes to
several seconds of latency during transient Wi-Fi interference, while
95\% of the NDT samples stay at or below 26.8\,ms, since the delay
distribution of \texttt{netem} is bounded. This makes the two
\emph{means} misleading to compare directly, since a handful of NET
outliers pull the mean upward. The \emph{medians} agree (22.3 vs.\
22.7\,ms), but the tails do not (95th percentile of 232.6 vs.\
26.8\,ms). Heavy-tailed, trace-driven, or AI-generated impairment
profiles (Section~\ref{sec:modeling}) are a natural extension to close
this gap.

\textbf{Virtual Queue Baseline Shift.} The minimum RTT of the NDT is
19.9\,ms, against about 2.0\,ms in NET. The Baseline profile gives
0.079\,ms, so OVS processing contributes little, and the floor comes
mainly from the delay injected by the profile on the AP virtual ports.
It is stable and predictable, so it can be reduced by lowering the delay
of the profile states.

\textbf{Packet Loss Distribution.} The loss distributions are close:
54\% of the NET and 51\% of the NDT sessions have loss of at most
0.5\%, and 30\% and 32\% have at least 10\%. The differences are at the
extremes. 19\% of the NET sessions have no loss, while the lowest NDT
loss is 0.29\%, because the NDT profile always injects some loss in its
low-loss state; and the highest loss is 67\% in NET and 40\% in NDT
(Fig.~\ref{fig:cdf}). Reproducing the loss-free periods and the bursts
would need a profile with a zero-loss state or a Gilbert-Elliott model
(Section~\ref{sec:modeling}).

\subsection{Limitations}
The results come from one case study and support the blueprint as a
proof of concept, not as a general validation, and as such, it has several limitations. First, the NET and NDT datasets were measured in different periods and are not paired. The NET channel varies from day to day (daily median RTT from 10.6 to 47.7\,ms), so a difference between the datasets cannot be attributed to the twin alone. Second, the NDT TCP results rest on 55 sessions, and the agreement of the mean packet loss depends on excluding five NDT sessions with 100\% UDP loss; with them included, the NDT mean would be 15.0\% (Table~\ref{tab:exclusions}). Third, the exclusion rules differ between the datasets because the causes differ, and each is listed in Table~\ref{tab:exclusions}. Fourth, the evaluated twin is measurement-calibrated: its profiles are set from earlier measurements, and the live recalibration described in Section~\ref{sec:modeling} is not evaluated here. Fifth, the evaluation does not test scale: the twin runs on one machine with six nodes (Table~\ref{tab:env}), E3 and E5 are single, short runs, and neither long-term operation nor larger topologies were tested. Finally, the evaluated twin is a behavioral emulation of a Wi-Fi edge cluster: it does not model channel contention, retransmissions, rate adaptation, or mobility, and the results do not extend to the full IoT-Edge-Cloud continuum.
\section{Conclusions and Future Research}
\label{sec:conclusion}
This tutorial showed how to build a lightweight, fully open-source NDT
from freely available tooling, how to model wireless impairments, and
how to assess twin fidelity to the physical network. The
blueprint was evaluated in one case study against a physical Edge
testbed, as a proof of concept, and the key takeaways for practitioners
building similar twins are provided next.

Four static Wi-Fi profiles show that the effects of a degraded Wi-Fi
channel (delay, jitter, loss, and a rate limit) can be reproduced at
the network layer with standard Linux tools. In a single run, the SDN
controller ONOS moved traffic to a secondary path in under one second,
and the latency on the affected path decreased from circa 101\,ms to
0.04\,ms. Priority rules kept latency-sensitive traffic apart from
best-effort traffic during congestion. Compared with the physical
network, the twin gives close results for median RTT and UDP
throughput over the multi-day NET archive.

Three gaps remain to be explored. First, TCP throughput in the twin is
lower than in the physical network (a median of 3.77 vs.\ 7.38\,Mbps).
The cause is not yet identified: the 500-byte segment size is used in
both datasets and does not explain the gap alone. Second, the
two-state profile does not reproduce the long periods of the physical channel without loss. Third, \texttt{netem} cannot produce the
occasional multi-second delay spikes of the physical Wi-Fi channel. The
last two need richer impairment models. The case study also showed a
pitfall to avoid: if the shaper's burst size is too small, TCP traffic
is starved, and its throughput drops to zero.

\subsection{Directions for Future Research}
\begin{figure}[htp!]
\centering\hyphenpenalty=10000\relax
\resizebox{\columnwidth}{!}{%
\begin{tikzpicture}[
  node distance=0.35cm and 0.9cm,
  g/.style={draw, rounded corners=2pt, align=center, text width=2.9cm,
            inner sep=3pt, font=\footnotesize, fill=gray!8},
  d/.style={draw, rounded corners=2pt, align=center, text width=3.1cm,
            inner sep=3pt, font=\footnotesize, fill=white},
  arr/.style={-{Stealth[length=5pt]}, thick}
]
\node[g] (g1) {TCP throughput below NET};
\node[g, below=of g1] (g2) {No long \mbox{loss-free} periods};
\node[g, below=of g2] (g3) {No multi-second delay spikes};
\node[d, right=of g1] (d1) {Find the TCP cause; size the shaper burst (or HTB/HFSC)};
\node[d, right=of g2] (d2) {Richer impairment models};
\node[d, right=of g3, dashed] (d3) {Beyond the gaps: automated recalibration};
\draw[arr] (g1) -- (d1);
\draw[arr] (g2) -- (d2);
\draw[arr] (g3.east) -- (d2.south west);
\node[font=\scriptsize\itshape, above=0.1cm of g1] {Gaps};
\node[font=\scriptsize\itshape, above=0.1cm of d1] {Directions};
\end{tikzpicture}}
\caption{Gaps found in the case study and the research directions that address them.}
\label{fig:future}
\end{figure}
 
Some directions for future research are summarized in Fig.~\ref{fig:future}. First, find the cause of the TCP gap: sizing the
shaper burst from the configured rate, or replacing TBF with HTB or
HFSC, would remove the need for the 500-byte segments and would show whether they explain it. Second, extend the validation by repeating the
measurements as paired, simultaneous runs on both networks with the
corrected shaper, by testing other topologies, such as multi-provider
and multi-tenant Edge-Cloud infrastructures, and by emulating the Wi-Fi
protocol itself. Third, automate recalibration: the live profile is designed to take its values periodically from current measurements, but this has not been evaluated in this paper. A recalibration loop, possibly with AI-generated profiles, could update delay, jitter, loss, and bandwidth as conditions change, so that the twin stays accurate without manual tuning.

\section*{Author Contributions}
\textbf{Josevany do Amaral:} Software development and lab setup (Containerlab topology
definitions, automated measurement framework); Investigation (all experimental runs, physical testbed, and NDT data collection); Visualization
(Grafana dashboards, result tables); Writing - initial draft.

\textbf{Rute C. Sofia:} Conceptualization (overall NDT
architecture and IRTF mapping; NDT
infrastructure and experimental design); Methodology
(benchmarking suite, Wi-Fi impairment profiles, data
collection pipeline); Formal Analysis (convergence analysis, divergence characterization,
artifact identification); Writing - final paper, analysis,  Review \& Editing; Supervision; Project
Administration; Acquisition of funding.
\section*{Acknowledgment}
 
This work has been developed in the context of the BayVFP RIVER project (fortiss-EnergyMiner) and in the context of the project SemComIIoT, funded by SGC, Grant agreement nr: M-0626.
\balance
\bibliographystyle{IEEEtran}
\bibliography{references}

\begin{IEEEbiography}
[{\includegraphics[width=1in,height=1.25in,clip,keepaspectratio]{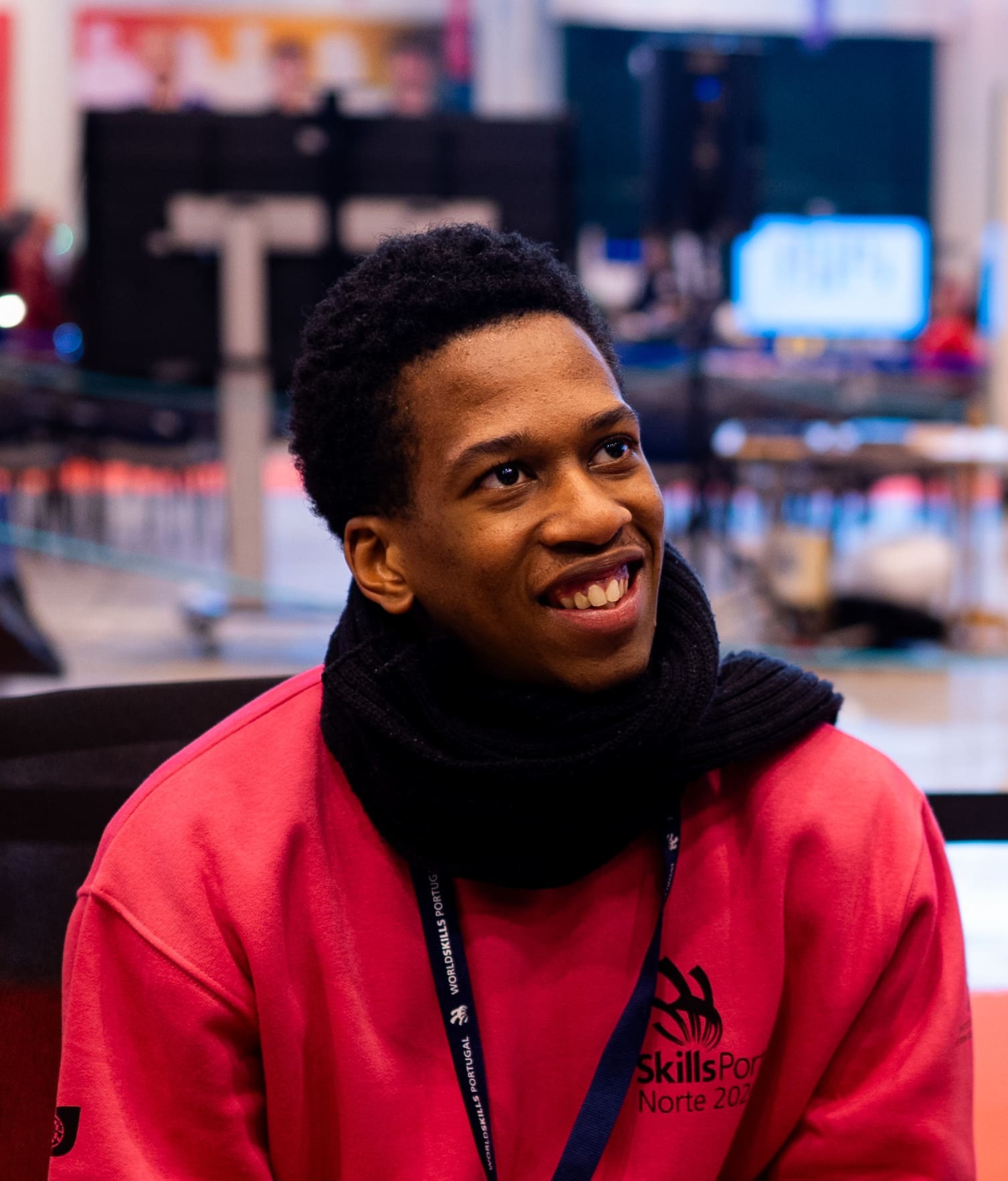}}]{\textbf{Josevany Y. F. do Amaral}} is a Mobile Development Technician (CTeSP, ISTEC – Instituto Superior de Tecnologias Avançadas, Lisbon), currently pursuing a B.Sc. in Computer Engineering at the same institution. His interests lie in software and mobile application development. He conducted the experimental work underlying this paper during an Erasmus+ research internship at Fortiss GmbH, Munich, where he developed and calibrated a Network Digital Twin for industrial IoT environments. He is a second-place winner at WorldSkills Norte 2024.

\end{IEEEbiography}

\begin{IEEEbiography}
[{\includegraphics[width=1in,height=1.25in,clip,keepaspectratio]{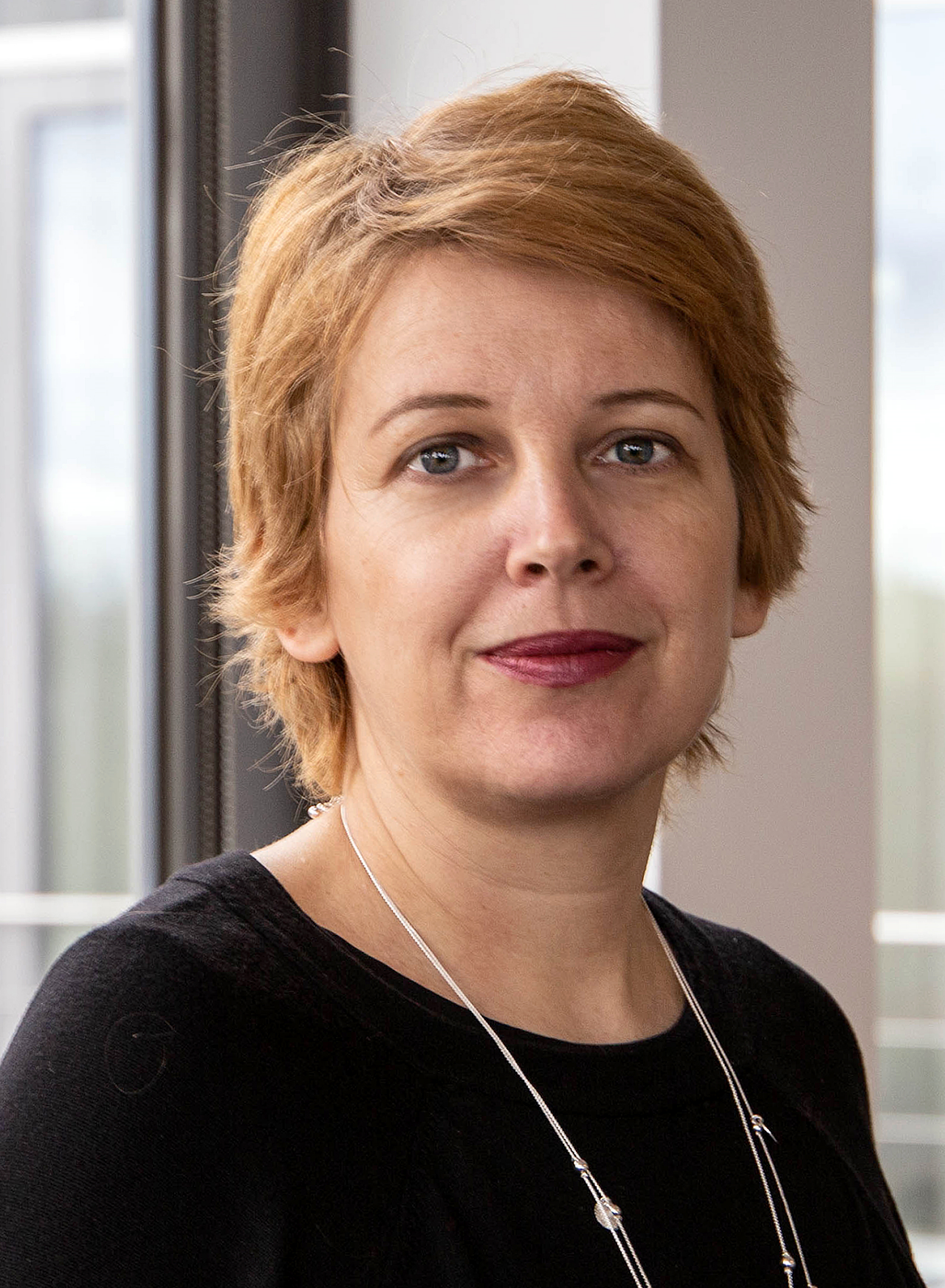}}]{Rute C. Sofia}
(Ph.D. 2004, IEEE Senior Member) leads the Industrial IoT competence field at fortiss, Munich, Germany, and is an Invited Associate Professor at Universidade Lus\'ofona. Her research interests include networking architectures and protocols, IoT, Edge computing, and AI for
networking. Earlier, she was a Senior Research Scientist at Universit\"at der Bundeswehr M\"unchen and at Siemens and Nokia Siemens
Networks, Munich; co-led the Internet Architectures and Networking area
at INESC TEC, Porto; and co-founded the COPELABS research unit and the
startup Senception, Lisbon. She has authored more than 90 peer-reviewed
publications and holds eight patents. She coordinates the CONASENSE 6G
platform, co-chairs the energy working group of AIOTI, and is a Board
Member of the Datacom Industry Association (DIA). She previously served
as ACM Europe Councilor and as IEEE ComSoc WICE Industrial Liaison
Deputy, and is an expert of the IEEE WIE network. She is an ACM Senior
Member and was recognized as Person of ACM in 2021.
\end{IEEEbiography}

\end{document}